\documentstyle[12pt,aaspp4]{article}

\newcommand{\de}{\delta}
\newcommand{\th}{\theta}

\newcommand{\si}{\sigma}
\newcommand{\bx}{{\bf x}}

\newcommand{\lan}{\langle}
\newcommand{\ran}{\rangle}

\newcommand{\be}{\begin{equation}}
\newcommand{\ee}{\end{equation}}
\newcommand{\bea}{\begin{eqnarray}}
\newcommand{\eea}{\end{eqnarray}}

\newcommand{\bef}{\begin{figure}}
\newcommand{\eef}{\end{figure}}

\def\spose#1{\hbox to 0pt{#1\hss}} 
\def\ltapprox{\mathrel{\spose{\lower 3pt\hbox{$\mathchar"218$}} 
 \raise 2.0pt\hbox{$\mathchar"13C$}}} 
\def\gtapprox{\mathrel{\spose{\lower 3pt\hbox{$\mathchar"218$}} 
 \raise 2.0pt\hbox{$\mathchar"13E$}}} 
\def\inapprox{\mathrel{\spose{\lower 3pt\hbox{$\mathchar"218$}} 
 \raise 2.0pt\hbox{$\mathchar"232$}}} 
 
\begin{document} 
 
\title{Bias and the power spectrum beyond the turn-over}
\author{Ruth Durrer \altaffilmark{1,2},
Andrea Gabrielli\altaffilmark{3,4,5},  
Michael Joyce \altaffilmark{6,7} and 
Francesco Sylos Labini\altaffilmark{8,9}}

\altaffiltext{1}{ruth.durrer@physics.unige.ch}

\altaffiltext{2}{D\'epartement de Physique Th\'eorique, 
Universit\'e de Gen\`eve,
24 quai Ernest Ansermet, CH-1211 Gen\`eve 4, Switzerland.}

\altaffiltext{3}{andrea@pil.phys.uniroma1.it}

\altaffiltext{4}{Dipartimento di Fisica, Universita' 
di Roma "La Sapienza", P.le Aldo Moro 2,  00185 Rome, Italy.} 

\altaffiltext{5}{Centro Studi e Ricerche "Enrico Fermi" e 
Museo Storico della Fisica, Via Panisperna 89 A, Compendio 
del Viminale, Palaz. F, 00184 Rome Italy.}

\altaffiltext{6}{joyce@lpnhep.in2p3.fr}

\altaffiltext{7}{Laboratoire de Physique Nucl\'eaire et de Hautes Energies,  
 Universit\'e de Paris VI, 4, Place Jussieu, 
Tour 33 -Rez de chaus\'ee, 75252 Paris Cedex 05, France.}

\altaffiltext{8}{francesco.sylos-labini@th.u-psud.fr}

\altaffiltext{9}{Laboratoire de  Physique Th\'eorique,
 Universit\'e Paris XI, B\^atiment 211, F-91405   
Orsay, France.}

\begin{abstract} 
Threshold biasing of a Gaussian random field 
gives a linear amplification of the reduced two point
correlation function at large distances. We show that
for standard cosmological models this {\it does not} translate 
into a linear amplification of the power spectrum (PS) 
neither at small $k$ not at large $k$. 
For standard CDM type
models 
the ``turn-over'' at small $k$ 
of the original PS disappears in the PS of the 
biased field for the 
physically relevant range of threshold parameters $\nu$.
In real space 
this difference is manifest in 
the asymptotic behaviour of the normalised mass variance in 
spheres of radius $R$,  which changes from 
the ``super-homogeneous'' behaviour 
$\sigma^2(R) \sim R^{-4}$ to a Poisson-like behaviour 
$\sigma^2_{\nu}(R) \sim R^{-3}$. This qualitative 
change results from the intrinsic stochasticity of the 
threshold sampling.  While our quantitative results
are specific to the simplest threshold biasing 
model, we argue that our qualitative conclusions should
be valid generically for any biasing mechanism involving
a scale-dependent amplification of the correlation
function. One implication is that the real-space 
correlation function will be  
a better instrument to probe for
the underlying Harrison Zeldovich spectrum
in the distribution of visible matter, as the 
characteristic asymptotic negative power-law 
$ \xi (r) \sim -r^{-4}$ tail is undistorted 
by biasing. 
\end{abstract}


\setcounter{footnote}{0}

The concept of bias has been introduced by Kaiser (1984),
primarily to explain the observed difference in amplitude
between the correlation function of galaxies and that
of galaxy clusters.
In this context
the underlying distribution of dark matter is treated 
as a correlated Gaussian density field.
The  galaxies
of different luminosities or galaxy clusters, are interpreted as
the peaks of the matter distribution, which have collapsed
by gravitational clustering. 
Different kind of objects are selected 
as peaks above a given threshold,
with a change in the threshold selecting
different regions of the underlying Gaussian field,
corresponding to fluctuations of differing amplitudes. 
The reduced two-point correlation 
function of the selected objects is then that 
of the peaks $  \xi_{\nu}(r)$, which is enhanced  
with respect to that of the underlying density field 
$\xi(r)$.
In a previous paper (Gabrielli, Sylos Labini \& Durrer, 2000 - GSLD00)  
some of us have discussed the problematic aspects of this 
mechanism. In particular the amplification of the 
correlation function is in fact only linear in the 
regime in which $\xi_{\nu}(r) \ll 1$ (see also Politzer \& Wise, 1984).
In the region of most observational relevance (where $\xi_\nu(r) \gg 1$) 
the correlation function is actually distorted at least exponentially.
Furthermore we have drawn attention
to the fact that the amplification of the correlation function 
by biasing reflects simply that
the distribution of peaks is {\it more clustered} 
because peaks are exponentially {\it sparser}.

In this letter we discuss a different aspect of this model
for bias. We are interested in understanding the effect
of biasing on the power spectrum (PS).
In particular we address here a qualitative change that
is caused to the matter perturbations in 
in standard cosmological models. In real
space this change manifests itself in a change from
sub-Poissonian behaviour of the mass variance at large
scales in the underlying density field, to Poissonian behaviour
of the same quantity for the biased field. In $k$ space 
this implies a distortion of the PS at small $k$. Our analysis
shows that this effect can be very important
observationally, as it can make the ``turn-over'' in 
the dark matter PS disappear from the 
PS of visible matter. Furthermore it shows 
the importance of measuring not just 
not just the PS of visible objects, but also their
real space correlation properties. Earlier works
about the effect of biasing in the PS 
can be found e.g. Coles (1993)  
and Scherrer \& Weinberg (1998).

Before considering threshold biasing, we recall
the relevant part of the analysis given in
Gabrielli, Joyce \& Sylos Labini (2002) 
(hereafter GJSL02). In this paper we have discussed the
meaning of the condition $P(0)=0$ satisfied by the 
PS  of all current standard type cosmological
models (with Harrison-Zeldovich like spectra $P(k) \sim k$
at small $k$). While this point
is often noted in the cosmological literature
(see e.g. Padmanabhan 1993), 
its significance and implications are not correctly
appreciated (see GJSL02 for discussion). 
It implies the requirement that the integral over all space of the
correlation function vanishes, meaning that in the
system there is an exact balance between
correlations and anti-correlations at all scales. 
This is a highly non-trivial, non-local, condition on the 
distribution.  Its specificity can be highlighted by
the following classification of all stationary stochastic 
processes into three categories: (i) For $P(0)= \infty$ the
fluctuations are like those in a critical long range
correlated system, (ii) for $P(0)= constant > 0$ the system is
Poisson-like at large scales e.g. any short-range 
positively correlated system such as a quasi-ideal gas at thermal 
equilibrium, and (iii) for $P(0)= 0$ the system
is what we have termed ``super-homogeneous''.
The reason for the use of this last term comes
from the fact that the three categories are
distinguished most strikingly in real space by 
the large distance behaviour of the 
mass variance in spheres, as one can show 
that
$P(0)=\lim_{V \rightarrow \infty} 
\frac{\langle (\Delta M (V))^2 \rangle}{\rho_o^2 V}
$
where $\langle (\Delta M (V))^2 \rangle$ is the mass
variance in a volume $V$ (and $\rho_o$ the mean mass density). 
In the Poisson type distribution this variance is 
proportional to the volume of the sphere, while
in the first category (critical systems) it grows more 
rapidly (with a limiting behaviour of the volume squared), while in 
the last (super-homogeneous distributions) the growth is slower 
than the Poissonian one.
In particular the case of the H-Z spectrum marks 
the transition to the limiting slowest possible
growth of this 
quantity for {\it any} stochastic distribution 
of points  (Beck 1987), which is a growth
proportional to the surface of the sphere. 

These super-homogeneous distributions are encountered
in various contexts in statistical physics. They
are described in this context as glass-like: they are 
highly ordered distributions like a lattice, but with 
full statistical isotropy and homogeneity. In  
Gabrielli et al. (2002) an example of a system
with such correlations at thermal equilibrium is given,
and a modification of this same system which should give
precisely the correlation of a standard cosmological model
is described. 

We now turn to the threshold biasing mechanism.
Following Kaiser (1984) we consider a stationary, 
isotropic and correlated continuous Gaussian 
random field,  $\de(\bx)$, with zero mean and
variance $\si^2=\lan \de(\bx)^2\ran$ in a volume $V$
as $V \rightarrow \infty$.
The marginal one-point probability density function of $\de$  is 
$ {\cal P}(\de) = {1\over \sqrt{2\pi}\si} e^{-{\de^2\over 2\si^2}} $.
Using ${\cal P}(\delta)$, we can calculate the  fraction of the 
volume $V$ with $\de(\bx)\ge \nu\si$,
$Q_1(\nu) = \int_{\nu\si}^\infty {\cal P}(\de) d\de \,.
$
The correlation function between the values of $\de(\bx)$ in two points
separated by a distance $r$ is given by 
$ \xi(r)=\lan \de(\bx)\de(\bx+ r{\bf n})\ran$. By definition, $ \xi(0)=\si^2$.
In this context, stationarity means that the variance, $\si^2$, and the
correlation function, $ \xi(r)$, do
not depend on $\bx$. Statistical isotropy means 
that $ \xi(r)$ does not depend on
the direction {\bf n}.
The goal is to determine the correlation function of 
local maxima from the
correlation function of the underlying density field. 
The problem can be simplified (Kaiser 1984) by computing the
correlation of {\it regions} above a certain threshold $\nu\sigma$
instead of the correlations of  {\it maxima}. However, these quantities are
closely related for values of $\nu$ significantly larger than 1.
We define the  threshold density, $\th_\nu(\bx)$ by
\be
 \th_\nu(\bx) \equiv \th(\de(\bx)-\nu\si)= \left\{ \begin{array}{ll}
        1 & \mbox{if }~~~ \de(\bx) \ge \nu\si \\
        0 & \mbox{else.}
\end{array} \right.
\ee
Note the qualitative difference between $\de$ which is a weighted
density field, and $\th_\nu$ which just defines uniform domains, all
having equal weight, and  $\langle \th_\nu(\bx) \rangle=Q_1$.

Let us now consider how the biasing changes the distribution
in relation to the classification we have given in terms
of $P(0)$. In what follows we show, for $\nu> \nu_o >0$,
($\nu_o$ given below)
\be
\label{result}
P_\nu(0) > P(0)
\ee
where $P_\nu(k)$ and $P(k)$ are the PS  
of the biased and underlying field respectively, i.e. the 
Fourier transform of $\xi_\nu(r)$ and of the normalised
underlying correlation function $\xi_c(r)\equiv \xi(r)/\sigma^2$ 
respectively. This result is independent of $\xi_c(r)$.
The correlation function $\xi_\nu(r)$ of the 
biased field is given (Kaiser 1984) by the expression 
\bea
 Q_1(\nu)^2 ( \xi_\nu(r) + 1) =  {1\over 2\pi\sqrt{1- \xi_c^2(r)}} 
        \int_\nu^\infty \int_\nu^\infty d\! \delta d\! \delta'
\nonumber  \\
\times \exp\left(-{(\delta^2+\delta'^2)-2 \xi_c(r)
\delta \delta' \over 2(1- \xi_c^2(r))}\right)\,
\label{xinuex} 
\eea
where the integrand on the right hand side is the two point
joint probability density for the Gaussian field 
Using the expression for $Q_1(\nu)$  given  above,
one can recast this after a simple change of variables into 
the form
\be
\label{xinu}
\xi_\nu(r)=\frac{\int_{\nu}^\infty dx e^{-x^2/2} 
\int_{\mu}^{\nu} dy e^{-y^2/2}}
{[\int_{\nu}^{\infty} dx e^{-x^2/2}]^2}. 
\ee
where $\mu=(\nu-\xi_cx)/\sqrt{1-\xi_c^2}$.
In this form it is evident that $\xi_c(r)=0 \Leftrightarrow \xi_\nu(r)=0$
and that $sign[\xi_\nu(r)]=sign[\xi_c(r)]$.

Taylor expanding this expression
about $\xi_c=0$,  we find\footnote{For the
expansion to all orders see Jensen \& Szalay (1986).}
\be
\xi_\nu(r)=b_1(\nu)\xi_c(r) + b_2(\nu)\xi_c^2(r)+...
\label{taylor}
\ee
with
\bea
b_1(\nu)=e^{-\nu^2/2}
\frac{\int_{\nu}^\infty dx x e^{-x^2/2}}
{[\int_{\nu}^{\infty} dx e^{-x^2/2}]^2} \\
b_2(\nu)=\frac{1}{2} \nu e^{-\nu^2/2}
\frac{\int_{\nu}^\infty dx (x^2-1) e^{-x^2/2}}
{[\int_{\nu}^{\infty} dx e^{-x^2/2}]^2}\,.
\eea
The first term gives the linear relation obtained by
Kaiser (1984), as $b_1(\nu) \approx \nu^2$ for $\nu \gg 1$,
valid in the regime $|\xi_c| \ll 1$ and $|\xi_\nu| \ll 1$.
It is easy to check that $b_2(\nu)$ is positive definite
for $\nu \ge 0$ (and $b_2(\nu)\approx \nu^4/2$ for $\nu \gg 1$), so
that to this order in $\xi_c(r)$ one has the bound 
\be
\xi_\nu(r) > b_1(\nu) \xi_c(r)  \; \; \mbox{for} \; \; \xi_\nu (r) \neq 0 
 \;\mbox{and} \; \nu > 0 \,.  
\label{bound} 
\ee 

If $|\xi_c(r)| \ll 1$ at all $r$ this bound suffices to give the desired
result (\ref{result}) for  all values of $\nu$ such that
$b_1(\nu) \geq 1$.  
As $b_1(\nu)$ is a monotonically increasing
function of $\nu$ (with $b_1(0)=2/\pi$) this is equivalent to the
requirement $\nu \geq \nu_o$ with $\nu_o$ such that $b_1(\nu_o)=1$
(i.e. $\nu_o\simeq 0.303$).
To show that there is a value of $\nu$ above which 
Eq.~(\ref{result}) is indeed satisfied for all 
 permitted values of $\xi_c$,
it suffices to find the threshold value $\nu_1\ge \nu_o$ such that
for all $\nu\ge \nu_1$ one has
\be
sign\left[\frac{d{\xi_\nu(r)}}{d{\xi_c(r)}} - b_1(\nu)\right] =
sign[\xi_c] \quad \forall r \;. 
\label{bound-conditions3} 
\ee
In fact this is a sufficient condition to have the exact curve 
$\xi_\nu(\xi_c)$, given by Eq.~(\ref{xinu}), 
all above the line $\xi_\nu=\xi_c$ for all $\xi_c$.
We have found numerically, using Eq. (\ref{xinu}), that 
this condition is satisfied  for $\nu_1 \simeq  0.38$.

Note that, if the condition (\ref{bound}) holds, this means simply 
that, relative to the asymptotic ($|\xi_c| \ll 1$ and $|\xi_\nu| \ll 1$)
linearly biased regime in which $\xi_\nu \approx b_1(\nu) \xi_c$, 
the anti-correlated regions are less amplified 
($|\xi_\nu| < b_1(\nu)|\xi_c|$) than the 
positively correlated regions ($|\xi_\nu| > b_1(\nu)|\xi_c|$). 
Thus the integral over the biased correlation function 
is always positive, and the bound (\ref{result}) thus holds.
Further it is easy to see that $P_\nu(0)$ is finite if
$P(0)$ is: $\xi_\nu$ is bounded for any value of $\nu$,
and, has the same convergence 
properties as $\xi_c$ at large distances. This implies
that, if the integral of $\xi_c$ over all space converges,
then also that of $\xi_\nu$ does.

In terms of the classification of distributions by $P(0)$
we thus draw the following conclusion: Both the critical 
type system (with $P(0)=\infty$) and Poisson type system
(with $P(0)=constant>0$) remain in the same class; the 
super-homogeneous distribution (with $P(0)=0$) however
becomes Poissonian ($P_\nu(0)=constant >0$). The essential
reason for these changes is simple: as discussed above 
the behaviour of the PS is the same as that 
of the mass variance at asymptotically large scales.
The biasing process is stochastic in nature, and introduces
a variance in the number of objects which is proportional
to the volume. This new variance will dominate asymptotically 
over that of the original distribution only if the latter is 
super-homogenous (i.e. its asymptotic normalised
variance is  sub-poissonian, decaying faster than Poisson). 
Consider for example the case of a perfect lattice, which is a
super-homogeneous distribution ($P(0)=0$) in which the 
normalised variance 
$\sigma^2(R)= \langle (\Delta M (R))^2 \rangle/\langle M(R)\rangle^2$
in a sphere of radius $R$ decays asymptotically 
as $1/R^4$. The distribution obtained 
by keeping (or rejecting) each point with probability 
$p$ (or $1-p$) is described by a simple binomial 
distribution, with a variance 
$\sigma^2(R) \propto p(1-p)/N \propto 1/R^{3}$
($N$ being the mean number of points inside a sphere).
Biasing is not  such a purely random sampling, 
but the effect of stochasticity 
as a source of Poisson variance at large scales 
is similar. Translated in terms of the PS it gives 
the result we have derived.

Now let us turn to the implications of this result
for cosmological models. Since such models have
$P(0)=0$ in the full matter spectrum, it is evident 
that we cannot have the behaviour 
$P_\nu(k) \propto b_1 (\nu)P(k)$ for small $k$ which
one might naively infer from the fact that
$\xi_\nu(r) \approx b_1(\nu) \xi_c(r)$ for large 
separations. Inevitably a non-linear distortion of
the biased PS at small $k$ relative to the underlying one
is induced. How important can the effect be qualitatively
for a realistic cosmological model?
To answer this question we consider the simple model PS
$P(k)=Ake^{-k/k_c}$. The differences with a cold dark
matter (CDM) model
- which has the same linear Harrison-Zeldovich form at 
small $k$ but a different (power-law) functional form
for large $k$ - are not fundamental here,  and this 
PS allows us to calculate the correlation function $\xi_c$ 
analytically (see GJSL02). This greatly simplifies our 
numerical calculation
of the biased PS $P_\nu(k)$, which we do by direct 
integration of $\xi_\nu(r)$ calculated using the
approximation 
\be
\xi_{\nu}(r)=\left[\sqrt{\frac{1+\xi_c(r)}{1-\xi_c(r)}}\exp\left(\nu^2
\frac{\xi_c}{1+\xi_c}\right)-1\right](1+o(\nu^{-1}))
\label{approx-xinu}
\ee
which is very accurate over most of the range of the integration
\footnote{This approximation is obtained by
expanding the full expression for $\xi_\nu(r)$ given
in Eq. (\ref{xinu}) in $1/\nu$, and further
assuming only
that $\nu \sqrt{(1-\xi_c)/(1+\xi_c)} \gg 1$. It is 
a much better approximation than that of 
Politzer and Wise both at small and larger values
of $\xi_c$. In particular it gives an asymptotic 
behaviour $\xi_\nu \approx (\nu^2+1) \xi_c$ for
$\nu^2\xi_c \ll 1$ which is a much better 
approximation to the exact behaviour at
typically relevant values of $\nu$ 
($b_1(1)\approx 2.4$)}.
In Figure \ref{fig1} we show $P_\nu(k)$
for various values of the threshold $\nu=1,2,3$. 
We see that the shape of the PS at small $k$ is
completely changed with respect to the underlying
PS. Indeed the main feature of the latter 
in this range - the display of a clear maximum 
and ``turn-over'' - is completely modified.
Qualitatively it is not difficult to understand
why this is so. The only characteristic scale
in the PS (and also in the correlation function)
is given by the turn-over (specified in our case 
by $k=k_{c}$). On the other hand, the value of $P_\nu (0)$ is 
just the integral over all space of $\xi_\nu$ 
which is proportional to the overall normalisation $A$ and (since
it is strictly positive) must be given on 
dimensional grounds by $Ak_c$ times some function 
which depends on $\nu$. For $\nu \simeq 1$ this 
function is of order one, so that 
$P_\nu (0) \sim max [P (k)]$. 

This last point is better illustrated  by considering
the integral 
$J_3(r,\nu)= 4 \pi \int_0^r x^2 \xi_\nu(x) dx 
$
which converges to $P_\nu (0) = \lim_{r \rightarrow \infty} J_3(r,\nu)$. 
In Figure \ref{fig2} the value obtained for it by
numerical integration of the {\it exact} expression 
given by Eq. (\ref{xinu}) for $\xi_\nu(r)$ is shown
for $\nu=1,2,3$.
We also show the same integral for $\xi_c$  which
converges to $P(0)=0$. While the latter 
decreases at large $r$, converging very slowly to 
zero (as $1/r$ since $\xi_c(r) \propto -1/r^4$ at large
scales), the former all converge towards a constant non-zero
value
We see that the integral
picks up its dominant contribution from scales
around (and above for $\nu=1$) $r \sim 10$ Mpc   
(see caption for explanation of the normalisations,
which are irrelevant for the present considerations).
From the inset in the figure, which shows both
$\xi_c(r)$ and $\xi_\nu(r)$, we see that this is
the scale below which the correlation function is
non-linearly amplified. Moreover it is shown that the smaller
scales at which $\xi_\nu(r)$ is most distorted 
relative to $\xi_c(r)$ do not contribute 
significantly to $J_3$ (because of the $r^2$ factor). 
This fact also explains the accuracy of the PS 
obtained using the approximation Eq. (\ref{approx-xinu})
for $\xi_\nu(r)$, which
can be seen by comparing the asymptotic values of
the integrals in Figure \ref{fig2} with $P_\nu(0)$
in Figure \ref{fig1}. Note that for $\nu=1$ the 
distortion away from linear is relatively weak in
the part of the correlation function which dominates
the integral in $J_3(r,\nu)$,  
and that there is even a non-negligible contribution
from the larger scales at which the correlation function
amplification is extremely close to linear.

Let us now draw our conclusions. We have calculated the
effect of the distortion at small $k$ of a biased PS 
relative to an underlying PS, which we have
shown to be an inevitable effect of biasing on 
cosmological models. 
In particular we have used a 
simplified model PS and the simplest (but reference)
threshold biasing scheme of Kaiser. The latter 
is not a realistic model of biasing for various reasons,
most notably because of the extremely strong 
non-linear distortion of the two-point correlation 
function at small scales whilst the observed ratio
of correlation between galaxies and clusters is 
approximately linear. Our results however should
be qualitatively correct for any biasing
scheme and standard CDM type model. The only
thing which we expect to depend on the biasing
scheme is the functional form of the PS 
to the right of the turn-over (for $k>k_c$), 
and correspondingly the shape of the correlation 
function at small scales, which will only make a
minor numerical change to our calculation.
The essential feature of biasing which 
brings about the effect we have discussed - 
a non-linear amplification of the 
correlation function - will be common to
any biasing model. In fact the converse of
what we have argued is that any biasing scheme
which is stochastic (i.e. any procedure for 
selecting sites for objects which is probabilistic)
must give such a non-linear distortion of the
correlation function for cosmological models: 
if the correlation function were amplified
linearly at all scales, the asymptotic behaviour of the 
variance will be sub-Poissonian instead of
Poissonian.

We finally draw two conclusions with respect to
the comparison of cosmological models with observational
data. In order to make the link to observations of galaxies or
clusters, current theories make use normally of
biasing in one form or another. Invariably however 
the measured PS from observations is fit to a model 
spectrum which is just a rescaled dark matter PS
(for a recent example see Lahav et al. 2002).
Our first conclusion is that such a behaviour cannot
be obtained by biasing 
as the PS is not linearly amplified 
neither at small nor at large
wavenumbers.
Our second conclusion is that it will be very useful 
to measure not just the PS at small $k$, but also
the real space correlation function at large
distances. Invariably only the first is considered
in analysis of observations of galaxy distributions
at large scales, because, it is argued, it is 
the natural probe given that cosmological models 
of structure formation are formulated in $k$ space.
We have seen however that biasing leads to a 
distortion of the underlying dark matter 
PS, while (insert panel of Figure \ref{fig2}) 
the correlation function at sufficiently large distance 
remains undistorted. In particular for standard models
with a H-Z PS at small $k$, the cleanest way to 
detect this behaviour in biased objects is 
by looking at the correlation function at large 
scales, which should maintain the 
behaviour $ \xi_{\nu}(r) \sim - r^{-4}$ associated
with the small $k$ behaviour of the underlying
dark matter PS. To address the viability
of measuring this behaviour in current and
forthcoming surveys of large scale structure 
requires in particular the detailed treatment 
of both the passage to the discrete distribution (we have
treated here always continuous density fields),
and the question of the variance of estimators
of $\xi(r)$. 

F.S.L. thanks the PostDoctoral support
of a Marie Curie Fellowship.
R.D. and F.S.L. acknowledge 
support of the Swiss National Science Foundation.
This work is supported by the TMR network FMRXCT980183  on Fractal
Structures  and Self-Organization.

\clearpage

\begin{figure}
\figcaption[f1.eps]{The PS $P_\nu (k)$ derived from the biased correlation
function $\xi_\nu(r)$ for values of the threshold $\nu=1,2,3$
is shown. The underlying correlation function which gives $\xi_c(r)$
is that derived from the $P(k)=Ake^{-k/k_c}$ and
the approximation given in Eq. (\ref{approx-xinu}) for
$\xi_\nu$ is used. The clear distortion
of the PS at small $k$ is seen, the ``turn-over'' 
in the underlying PS essentially disappears
already for $\nu=1$.  
The constants $k_c$ and $A$ are fixed by $\xi_c(0)=1$ and the 
requirement that $\xi_c(r)=0$ at $r=38$ Mpc. 
The latter is taken as in a typical CDM model
(see e.g. Padmanabhan 1993).
We could alternatively fix the wavenumber at the 
maximum of the PS.
{\label{fig1}}}
\end{figure}

\begin{figure}
\figcaption[f2.eps]{The integral $J_3(r)$ for the same 
underlying correlation function as in 
Figure \ref{fig1} and for the same range of
values of $\nu$, calculated with the exact expression
Eq. (\ref{xinu}) for $\xi_\nu(r)$. 
Also shown is the analagous integral of $\xi_c$. While 
the latter converges slowly to its asymptotic value
of $P(0)=0$, the other integrals converge to 
constant non-zero $P_\nu(0)$. They are dominated by the range 
$r \sim 10$ Mpc where, as can be seen from the inset which 
shows both $\xi_c$ and each of the $\xi_\nu(r)$,
the correlation functions
$\xi_\nu(r)$ are amplified non-linearly. The contribution
from the extremely amplified region at small $r$ is
small (because of the $r^2$ factor in the integral),
which also makes the approximation in calculating the PS
with Eq. (\ref{approx-xinu})
very accurate, as can be checked by comparing the asymptotic 
values with those of $P_\nu(0)$ in Figure \ref{fig1}.
{\label{fig2}}}
\end{figure}

\end{document}